\documentclass[aps,pra, nopacs,twocolumn,twoside,floatfix,a4,superscriptaddress]{revtex4}
\usepackage{hyperref}
\usepackage{bbm}
\usepackage{amsmath, amssymb}
\usepackage{color}
\usepackage{graphicx,epsfig}
\usepackage{times} 

\begin{document}

\title{Stronger Quantum Correlations with Loophole-free Post-selection}

\author{Matty J. Hoban}
\email{m.hoban@ucl.ac.uk}
\affiliation{Department of Physics and Astronomy, University College London, Gower Street, London WC1E 6BT, United Kingdom.}
\author{Dan E. Browne}
\affiliation{Department of Physics and Astronomy, University College London, Gower Street, London WC1E 6BT, United Kingdom.}

\date{\today}
\begin{abstract}
\noindent
One of the most striking non-classical features of quantum mechanics is in the correlations it predicts between spatially separated measurements.  In local hidden variable theories, correlations are constrained by Bell inequalities, but quantum correlations violate these. However, experimental imperfections lead to ``loopholes''  whereby LHV correlations are no longer constrained by Bell inequalities, and violations can be described by LHV theories. For example, loopholes can emerge through selective detection of
events. In this letter, we introduce a clean, operational picture of multi-party Bell tests, and show that there exists a non-trivial form of loophole-free post-selection. Surprisingly, the same post-selection can enhance quantum correlations, and unlock a connection between non-classical correlations and non-classical computation.
\end{abstract}
\maketitle
\noindent
\noindent
The correlations in classical physics, or more generally, local hidden variable (LHV) theories are famously constrained by the Bell inequalities \cite{bell}; even more notably, these are violated by quantum correlations. Characterising this quantum violation remains an open problem. Recently it has been proposed that quantum correlations are characterised by some principle, at least in the bipartite setting \cite{unc,mac,cc,ic}. However, for multi-party correlations an answer remains uncertain even though interesting particular examples do exist \cite{gyni}.

In addition, despite increasingly sophisticated experiments, loopholes allow LHV theories to simulate quantum correlations through various experimental imperfections \cite{loophole}. In the presence of imperfect detectors, the fact that detected events are not a fair sample of the actual set of events results in the detection loophole \cite{loophole}. Photonic tests of Bell inequalities \cite{zeilinger} currently suffer from this detection loophole. Despite avoiding the detection loophole, ion trap tests \cite{rowe} cannot yet achieve necessary space-like separation. However, photon-mediated ion entanglement \cite{monroe}, and progress in the efficiency of creating and detecting photons, mean that a loophole-free Bell test may not be far away. Until then, understanding the effect of post-selection is key to progress. 

\begin{figure}
	\centering
		\includegraphics[width=0.33\textwidth]{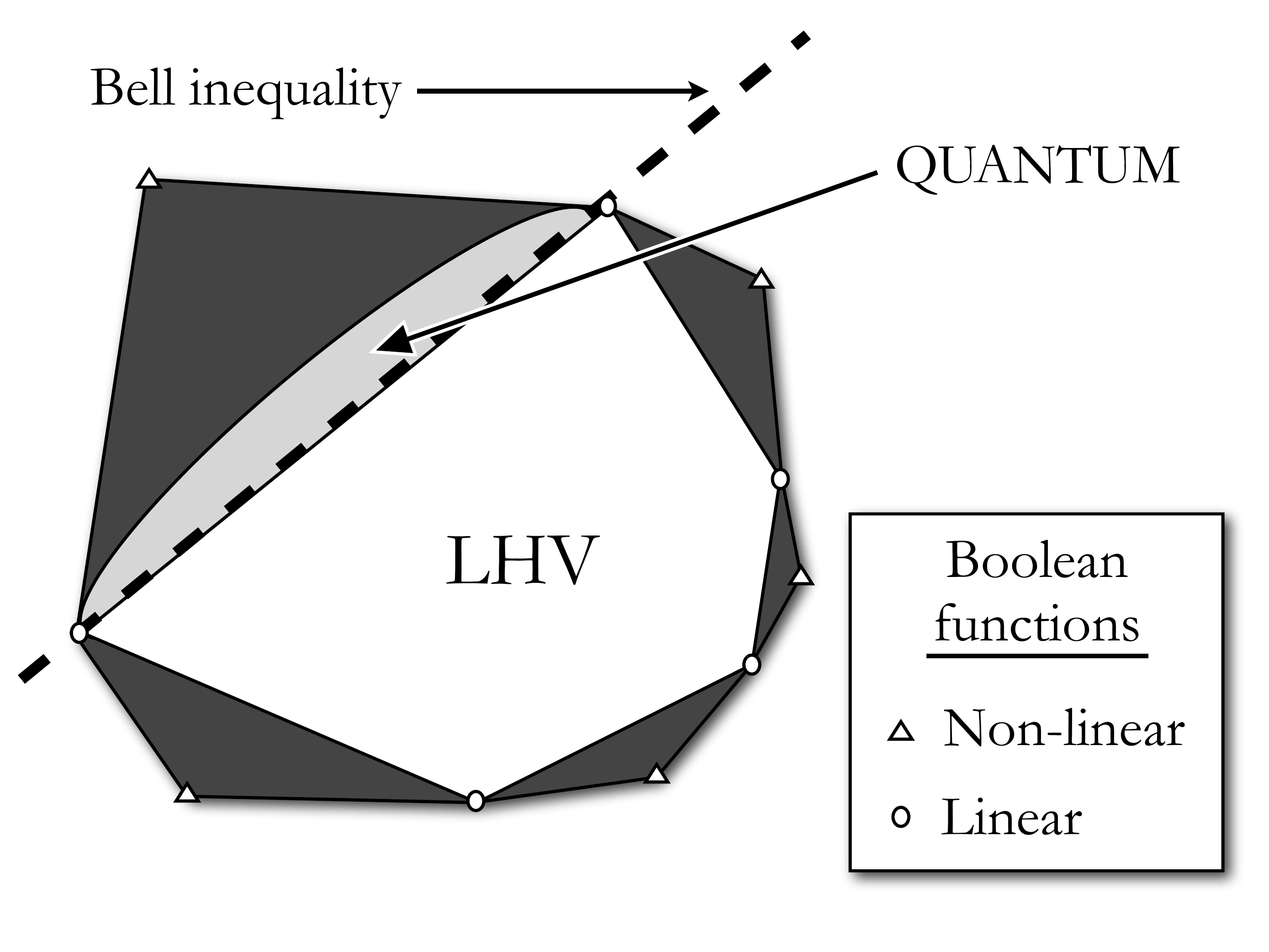}
			\caption{The correlators we study in a multi-party CHSH experiments can be thought of as stochastic maps. The full space of correlators in $\mathbb{R}^{2^n}$ is then the convex hull of deterministic maps, each of which is a \emph{linear} Boolean function, labeled by schematically by circles (see Theorem 1). The facet-defining Bell inequalities correspond to facets of this LHV polytope. The quantum correlators cover a larger region, corresponding to Bell inequality violations.}
			\label{fig:fig1}
\end{figure}

An elegant and powerful approach to the study of Bell inequalities has been through a geometrical point-of-view \cite{pitowsky, froissart}. As we shall introduce in detail below, a correlation can be represented by a vector in real space. The set of correlations achieved in any theory are then defined in a particular `region' of this real space. For example, in an LHV theory this region is a polytope whose facets are Bell inequalities \cite{pitowsky, froissart, ww, zb}. This is schematically described in fig. \ref{fig:fig1} where different theories form a hierarchy of regions in this correlation space and Bell inequalities bound the region of LHV correlations.

In this letter, we generalise the construction of multi-party Clauser-Horne-Shimony-Holt (CHSH) Bell inequality tests \cite{chsh, ww,zb} by introducing a new way of analysing the data in these experiments. Specifically, we introduce two methods of data selection, which we call ``setting post-selection'' (SP) and setting-output post-selection (SOP). We show that these methods allow one to derive new families of Bell inequalities, and, in particular  when these conditions which establish the post-selection are constrained to be \emph{linear}, the sets of correlations for LHV theories remains the same as in the standard CHSH framework. 
However, we also show that the set of quantum correlations is not invariant under these forms of post-selection leading to larger violations of Bell inequalities. 
What is more, the post-selection described in this letter also allows for the adaptive measurements in Measurement-based Quantum Computation \cite{mbqc} to be simulated in a CHSH-like test, thus making for the first time, a concrete connection between the correlations arising in this model and Bell inequality violation.


Let us now define what precisely we mean by a ``multi-party CHSH experiment'' \cite{ww,zb}. This is an experiment with $n$ parties each of which makes a measurement which has two settings and two outcomes. In these experiments,  data is collected under the following assumptions: \emph{Assumption 1}, all measurements made are space-like separated; \emph{Assumption 2}, the choice of measurement is uniformly random and private (known as measurement independence or the `free will' assumption). 

 Let $s$ be an $n$-bit string which represents the measurement settings; the $j$th party's measurement choice is labelled by the bit $s_{j}\in\{0,1\}$; measurement outcomes for each $s_{j}$ are $(-1)^{m_{j}}$, labelled by the bit $m_{j}\in\{0,1\}$. Note that we write the $j$th bit of bit-string $y$ as $y_j$. Thus measurement settings and outcomes for all parties are $n$-length bit-strings $s$ and $m$ respectively. In a CHSH experiment one studies the expectation values of the joint outcome $(-1)^{\sum_{j}^{n}m_{j}}$ for each choice of measurements $s$. These can be equivalently, and more conveniently, expressed as conditional probabilities (correlators) $p(\bigoplus_{j=1}^{n}m_{j}=1|s)$ where $\bigoplus$ represents addition modulo $2$.

The correlators $p(\bigoplus_{j=1}^{n}m_{j}=1|s)$ are a set of $2^n$ real numbers between 0 and 1. Hence each set of experimental data can be represented by a $2^{n}$-dimension vector $\vec{p}$ inhabiting a hypercube, whose vertices $\vec{p}_{f}$ are the set of binary vectors, with elements $0$ and $1$. There is a one-to-one correspondence between each of the extremal vectors $\vec{p}_{f}$ and the set of Boolean functions, i.e.  $p_{f}(\bigoplus_{j=1}^{n}m_{j}=1|s)=f(s)$ where $f(s)\in\{0,1\}$. This reflects the well-known correspondence between conditional probabilities and stochastic (or in the case of the extreme points deterministic) maps. 

\textit{LHV correlators} - The convexity of classical probability theory implies that regions of LHV correlators are always described by convex polytopes \cite{froissart, pitowsky,fine}. While the region of quantum correlators is also convex, it is not polytopic \cite{pitowsky}. We now present a very simple and compact result describing the region of LHV correlators in terms of the stochastic maps $f(s)$ that the correlators achieve in our model.
\newline

\noindent
\textbf{Theorem 1} \textit{In a multi-party CHSH experiment with settings $s$ and outputs $m$, all correlators $p(\bigoplus_{j=1}^{n}m_{j}=1|s)$ from a LHV theory lie within the convex hull of deterministic linear Boolean functions $f(s)$ on $s$.}

\textit{Proof} Firstly, we shall consider the deterministic maps achievable in a LHV  theory. As Fine has shown \cite{fine}, the full region of correlators will be the convex hull of these.  Due to the assumptions of locality and measurement independence in LHV theories, each outcome $m_{j}$ is only dependent on $s_{j}$ and $\lambda$, a variable shared by all parties which is necessarily independent of $s$. Without loss of generality, we can write $m_{j}=a_{j}(\lambda)s_{j}\oplus b_{j}(\lambda)$ where $a_{j}(\lambda), b_{j}(\lambda)\in\{0,1\}$ are dependent only on $\lambda$. Note that these  are the four one-bit Boolean functions, and each is (trivially) linear in $s_j$. The deterministic $n$-party correlators then are obtained from $\bigoplus_{j=1}^{n}m_{j}= \bigoplus_{j=1}^{n}a_{j}(\lambda)s_{j} \oplus b(\lambda)$ where $b(\lambda)=\bigoplus_{j=1}^{n}b_{j}$. These are all linear in $s$ and indeed  all $n$-bit linear functions are represented. Thus the LHV polytope is the convex hull of all $2^{n+1}$ linear Boolean functions on $s$. $\square$
\newline

The region of LHV correlators identified in Theorem 1 is equivalent to that characterised by Werner and Wolf \cite{ww} and independently by \.{Z}ukowski and Brukner \cite{zb}. A simple property of Boolean functions, \emph{linearity}, thus characterises the LHV correlators in our experiment. In addition to its vertex description, a convex polytope can be defined as the intersection of the half-spaces specified by some linear inequalities. The linear inequalities defining the facets of the polytope of LHV correlators are nothing other than the facet-defining Bell inequalities. The computation of these inequalities from vertices is computationally hard \cite{pitowsky}. A key advantage of our approach is that one can prove some general results without the need for facet Bell inequalities. For example, we see immediately that correlators which can \emph{only} be written as a convex combination with one or more non-linear functions must lie outside the LHV polytope.

In standard Bell inequality tests, loopholes arise when some aspect of the experiment allow correlations in an LHV theory which violate a Bell inequality. Theorem 1 tells us that loopholes in CHSH experiments can be understood in a simple way. All correlators which lie outside the LHV region must necessarily contain an admixture of a non-linear map. We can thus associate loopholes in CHSH experiments as sources of non-linearity. We define any modification of the standard CHSH experiment after which the region of LHV correlators remains inside the convex hull of linear functions \emph{loophole-free}.

Post-selection in an experiment is the rejection of a proportion of experimental data according to certain criteria. In this paper, we shall use the term in a slightly more general sense (made explicit below) encompassing both the rejection \emph{and relabelling} of experimental data. In general, post-selection is not loophole free. The detection loophole can be arise from \emph{post-selection} on measurement data now illustrated with a two-party example, modified from \cite{berry}. Consider this specific LHV model: the first and second parties' output are $m_1=s_1\oplus t$ and $m_{2}=t s_2$ respectively, where $t\in\{0,1\}$ is random. We now post-select on data satisfying $m_1=0$. This maps $s_1$ onto $t$, and results in $m_2=s_1s_2$. This function is clearly non-linear, and the associated correlator violates a Bell inequality. This example illustrates one way in which post-selection causes loopholes. It can allow one or more input bits to be mapped onto the shared hidden variables (cf. \cite{gisin}), transmitting data to other parties. The full detection loophole can be understood in a similar way, with a variation on the above model saturating the upper bounds derived by Garg and Mermin \cite{loophole, thesis}. While post-selection can lead to loopholes, we want to incorporate post-selection at a more fundamental level and avoid loopholes.

In constructing the correlators $p(\bigoplus_{j=1}^{n}m_{j}=1|s)$ as conditional probabilities, the bit-string $s$ has two roles: firstly, it describes the conditioning for the conditional probabilities, i.e. the stochastic map performing the function $f(s)$ is conditioned upon $s$; and secondly, $s$ specifies the measurement settings. We shall now reformulate and generalise this by separating these two aspects. We introduce a new bit-string $x$ that takes on the first role of conditioning, where the size $|x|$ of $x$ satisfies $|x|\leq n$. We then study correlators $p(\bigoplus_{j=1}^{n}m_{j}=1|x)$ and give them meaning by fixing a relationship between $x$, $m$ and $s$. 

In this framework one can recover the standard CHSH experiment (i.e. Theorem 1) by setting $x=s$, one can also perform relabellings of measurement settings by setting $s_{j}=g_{j}(x_{j})$ where $g_{j}(x)$ is a Boolean function. Going further, one can introduce data rejection into this formalism by setting $|x|<n$.

We shall now focus on two particular families of post-selection strategies and, in each case, identify the full region of correlators $p(\bigoplus_{j=1}^{n}m_{j}=1|x)$ achievable in a LHV theory. The first of these is setting-postselection (SP). In SP, we fix $s$ as a function of $x$ alone, i.e. to compute $p(\bigoplus_{j=1}^{n}m_{j}=1|x)$ we consider the statistics of $m$ on data where the settings $s_j$ are each equal to a function $g_j(x)$. If we make the further restriction that $g_j(x)$ be linear in $x$ we find that such post-selection is loophole-free. Stated more precisely:
\newline

\noindent
\textbf{Theorem 2} \textit{In a multi-party CHSH experiment with settings $s$ and outputs $m$, all correlators $p(\bigoplus_{j=1}^{n}m_{j}=1|x)$ from a LHV theory, after post-selecting on settings $s_{j}=g_{j}(x)$, lie within the convex hull of linear functions $f(x)$ on $x$ iff all $g_{j}(x)$ are linear.}

\textit{Proof} We follow the same strategy as the proof of theorem 1. Since every $s_j$ is a linear function of $x$, every $m_j$ remains a linear function of $x$. The remainder of the proof is identical to the above.$\square$
\newline

This theorem tells us that the polytope of linear functions does not just define the traditional Bell inequalities, but in fact a much broader range. In fact, starting from any standard CHSH-type inequality, any linear relabelling of the measurement settings will give an equally valid Bell inequality. For example, if $|x|=2$, one of the facet inequalities will be the standard CHSH inequality. By theorem 2, this inequality immediately implies the GHZ-Mermin inequality \cite{ghz, mermin} via the SP post-selection on $s_{1}=x_{1}$, $s_{2}=x_{2}$ and $s_{3}=x_{1}\oplus x_{2}$.

SP post-selection introduces dependences between measurement settings, and thus, at first sight it may seem surprising that the LHV region maintains its structure. In some sense this may be seen as a reduction in the ``freedom of choice" in measurements. The effect of reduction of free-choice in Bell experiments has been studied elsewhere \cite{gisin} and shown, in general, to lead to loopholes. We must make a distinction here, the correlations in settings introduced by SP post-selection are very special, leading to linear relationships between them. In this special case, theorem 2 tells us that this reduction in setting independence is, in contrast, loophole-free.

Now we show there is a post-selection that not only fixes a relationship between settings but \emph{also measurement outcomes} can be implemented, again without altering this structure. Such post-selection, which we call setting-output post-selection (SOP), allows us to simulate adaptive measurements, and also allows us to simulate signalling correlations. Again, if these relationships are constrained to being linear, the LHV region maintains its classic form, as made precise in this following theorem.
\newline


\noindent
\textbf{Theorem 3} \textit{In a multi-party CHSH experiment with settings $s$ and outputs $m$, all correlators $p(\bigoplus_{j=1}^{n}m_{j}=1|x)$ from a LHV theory, after post-selecting on settings $s_{j}=g_{j}({m^{\setminus j}},x)$ (where $m^{\setminus j} =m\!\setminus\! m_{j} $ is bit-string $m$ with the element $m_{j}$ removed),  lie within the convex hull of linear functions $f(x)$ on $x$ iff all $g_{j}(m^{\setminus j},x)$ are linear.}

\textit{Proof} We may rewrite $g_{j}(m^{\setminus j},x)$ as $g^{(1)}_{j}(m^{\setminus j})\oplus g^{(2)}_{j}(x)$ where $g^{(2)}_{j}(x)$ and $g^{(1)}_{j}(m^{\setminus j})$ are both linear functions depending on $x$ and $m^{\setminus j}$ respectively. Considering each site's deterministic map we obtain $m_{j}\oplus a_{j}(\lambda)g^{(1)}_{j}(m^{\setminus j})=a_{j}(\lambda)g^{(2)}_{j}(x)\oplus b_{j}(\lambda)$. We can see that (firstly by assuming that $\lambda$ is independent of $x$) all deterministic maps must be linear functions on $x$.


If the post-selection results in $\lambda$ being correlated to $x$, then it is possible to achieve non-linear functions through values of $a_{j}(\lambda)g^{(2)}_{j}(x)$ in $m_{j}$. We now show that $a_{j}(\lambda)$ always remains independent of $x$. The outcomes in $m^{\setminus j}$ contain information about $\lambda$, but $s_{j}$ is random and uncorrelated to $\lambda$, $m$ and $x$. Therefore $g^{(1)}_{j}(m^{\setminus j})=g^{(2)}_{j}(x)\oplus s_{j}$ means that $g^{(1)}_{j}(m^{\setminus j})$ is random and uncorrelated to $g^{(2)}_{j}(x)$ \footnote{If $g^{(2)}_{j}(x)=0$, the bit $s_j$ does become correlated with other sites' measurements $m_k$ and hence $\lambda$ but $s_j$ will be uncorrelated to $x$. If $g^{(1)}_{j}(m^{\setminus j})=0$, we recover SP.}. These random bits $s_j$ play the role of the pad-bit in one-time pad cryptography which Shannon \cite{shannon} proved is perfectly secure for encrypting messages.

If $g_{j}(m^{\setminus j},x)$ becomes non-linear then we can see as before that the stochastic map can always be this function $f(x)=g_{j}(m^{\setminus j},x)$. Since values of $m^{\setminus j}$ can be made to be equal to values of $x$, there always exists a non-linear function in $x$ if $g_{j}(m^{\setminus j},x)$ is non-linear. $\square$
\newline

\textit{Quantum correlations under post-selection} We have seen that linear SP or SOP does not change the structure of the correlations in LHV theories. This may seem surprising given that we can simulate adaptive-measurements. Indeed, this is a special property of the LHV correlations. The set of quantum correlators is a counter-example. It is not invariant. Indeed the set of quantum correlators can (for fixed $|x|$) be larger under linear SOP than under linear SP. 
\newline

\noindent
\textbf{Theorem 4} \textit{In a multi-party CHSH experiment with fixed $|x|$ and $n$, there exists values of $|x|$ and $n$ for which the region of quantum correlators $p(\bigoplus_{j=1}^{n}m_{j}=1|x)$ under linear SOP post-selection, is strictly larger than the the region of quantum correlators under linear SP post-selection.}

\begin{figure}
	\centering
		\includegraphics[width=0.35\textwidth]{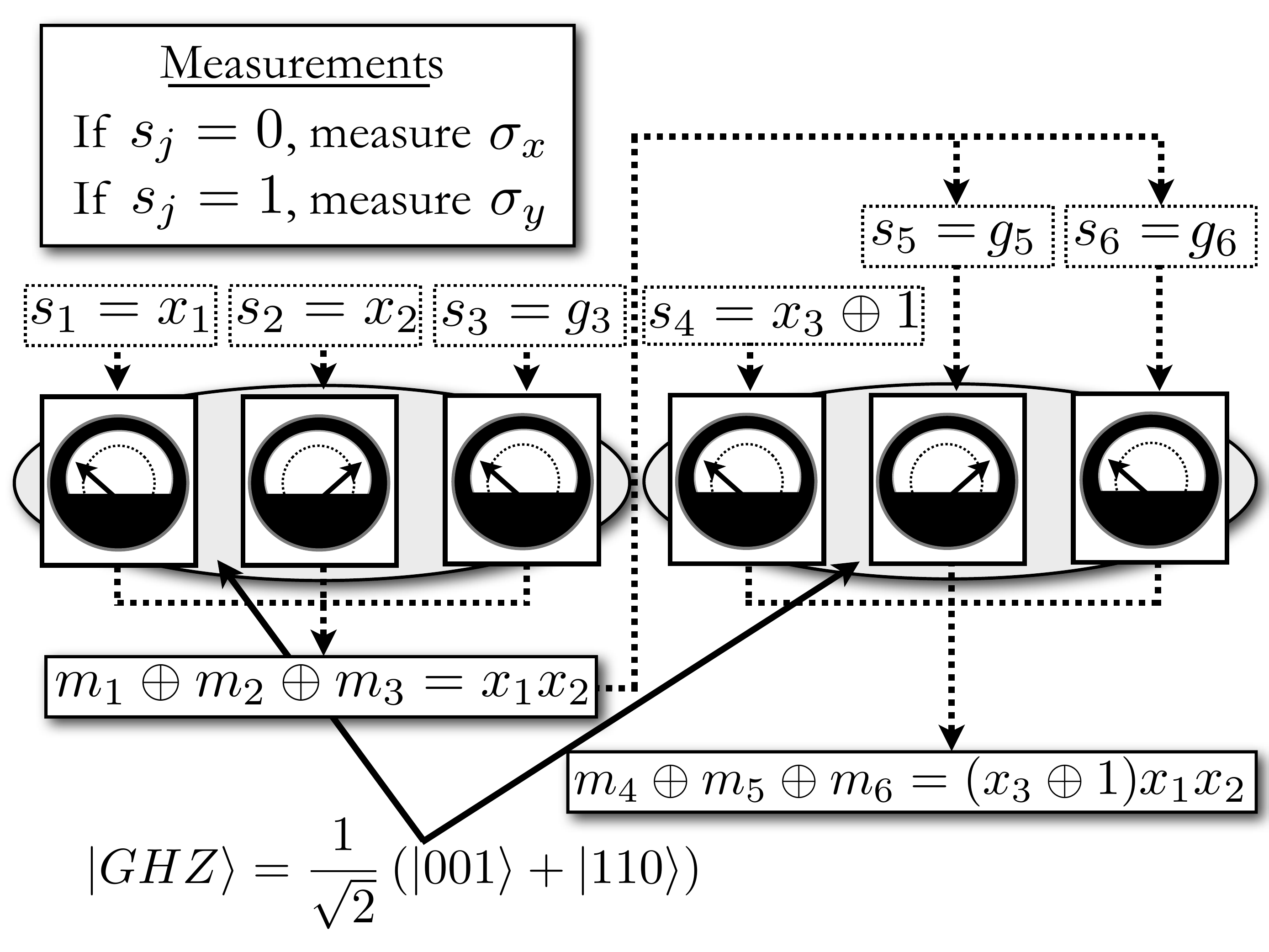}
			\caption{We can achieve $\bigoplus_{j=1}^{6}m_{j}=x_{1}x_{2}x_{3}$ with SOP for $n=6$ parties. The first three parties on the left and the other three parties on the right each share $|GHZ\rangle$, a GHZ state. Each party makes measurements corresponding to the $s_{j}$ described. For three parties sharing this GHZ state $m_{i}\oplus m_{j}\oplus m_{k}=s_{i}s_{j}$ if $s_{k}=s_{i}\oplus s_{j}$ \cite{anders}. Therefore we post-select on values of $s_{j}$ as shown in the dashed boxes where $g_{3}=x_{1}\oplus x_{2}$, $g_{5}=m_{1}\oplus m_{2}\oplus m_{3}$ and $g_{6}=m_{1}\oplus m_{2}\oplus m_{3}\oplus x_{3}\oplus 1$. This dependency on measurement outcomes is represented by the dashed lines showing \emph{flow} of information; this results in the non-linear function described.}
			\label{fig:fig2}
\end{figure}

\textit{Proof} We prove this by example. Consider $n=6$ parties and $x$ with length $|x|=3$. The quantum correlator $p(\bigoplus_{j=1}^{6}m_{j}=1|x)=x_{1}x_{2}x_{3}$ for all $x$ \emph{cannot} be achieved with only SP as shown in \cite{hoban}. However, this can be achieved with the adaptivity incorporated into SOP as shown in fig. \ref{fig:fig2}. We use the fact that the function $f(x)=x_{i}x_{j}$ can be performed deterministically for $3$ parties with a Greenberger-Horne-Zeilinger (GHZ) state \cite{ghz,anders}. Put simply, SOP composes these functions adaptively to perform $x_{1}x_{2}x_{3}$. $\square$
\newline

This simple argument shows that SOP may enlarge the region of quantum correlators for $|x|=3$ and $n=6$, and straightforward modification of this example for other monomial functions will generate examples for larger $|x|$ and $n$. If $n$ is unbounded, then the quantum correlators can access the entire correlator hypercube (i.e. the quantum correlations are maximally non-local) under SP, and hence also SOP. This is an implication of the results in \cite{hoban}. It is natural to ask whether this enhancement there an enhancement for simpler experiments.  As we show in the supplementary section \cite{supplement} the experimentally feasible $n=2$ setting there is \emph{no enhancement} from SOP, making this effect a truly multi-party phenomenon.

\textit{Summary and outlook} - We have shown a simple and operational picture of LHV correlators in terms of computational expressiveness for a particular multi-party Bell-type experiment. Some of the methods here can be extended here to the scenario with more settings and outcomes but we need to move beyond simple Boolean functions. We can also extend the loophole-free post-selection proof to considering Bell inequalities involving marginals as well as correlators such as the Clauser-Horne inequality \cite{ch}. 

We can enlarge the region of quantum correlators. However, the extent of this enlargement needs further investigation. The region of quantum correlators without post-selection can be characterised by a simple variational expression \cite{ww}, but this approach fails with SOP. A full characterisation of the SOP-enhanced quantum region remains an open question. The result motivates many other open questions, e.g. are there forms of post-selection that are not completely loophole-free but still enlarge quantum correlators?

One might naturally ask whether these results can be used to close a detection loophole? This is not actually possible with the methods presented in this work. The difficulty lies in the fact that the detection loophole fundamentally introduces the non-linear correlators described earlier; this irreducibly changes the structure of the LHV correlator space. Just as with the $6$-site example shown, SOP can ``amplify'' non-linearity which could perhaps also enhance a detection loophole since it is also associated with non-linearity.

Despite the current limited utility of this work experimentally we hope this work motivates new directions and insights into programmes for axiomatising quantum correlators \cite{unc,mac,cc,ic}. Since LHV theories are not affected by our post-selection, we hope that a principle that captures quantum mechanics can also capture the SOP enlarged quantum correlators. More generally, we hope it motivates a new appreciation of the novelty of multi-party quantum effects.

There is an important connection between our post-selection model and Measurement-based quantum computation (MBQC) \cite{mbqc}. Strikingly, SOP, is precisely the post-selection needed to simulate the adaptive measurements in Raussendorf and Briegel's cluster state model \cite{mbqc}.  Thus our method will allow, for the first time, the full MBQC model to be studied using Bell inequalities, which may reveal the fundamental physics behind it. Is it even possible that quantum correlators can be characterised by a computational principle (cf. \cite{barrett})?

The Bell inequalities have been studied for over 40 years, and continue to throw up new surprises. We hope this work contributes further to our understanding of what separates the quantum from the classical.

\textit{Acknowledgements} - We thank Joel Wallman for very useful comments and Julien Degorre, Daniel Oi, Janet Anders, Matt Leifer and Peter Janotta for interesting discussions. MJH acknowledges financial support from EPSRC and DEB is supported by the Leverhulme Trust.


\begin{thebibliography}{99}
\bibitem{bell} J. S. Bell, \textit{Physics 1, 195 (1964).}
\bibitem{ic} M. Pawlowski et al, \textit{Nature 461, 1101 (2009).}
\bibitem{mac} M. Navascues, H. Wunderlich, \textit{Proc. Roy. Soc. Lond. A 466, 881-890 (2009).}
\bibitem{cc} G. Brassard et al, \textit{Phys. Rev. Lett. 96, 250401, (2006).}
\bibitem{unc} J. Oppenheim, S. Wehner, \textit{Science 330, 6007, 1072-1074 (2010).}
\bibitem{gyni} M. L. Almeida et al, \textit{Phys. Rev. Lett. 104, 230404 (2010).}
\bibitem{loophole} P. Pearle, \textit{Phys. Rev. D, 2, 1418-25 (1970)}; A. Garg and N. D. Mermin, \textit{Phys. Rev. D 35, 3831 - 3835 (1987).}
\bibitem{zeilinger} G. Weihs, T. Jennewein, C. Simon, H. Weinfurter and A. Zeilinger, \textit{Phys. Rev. Lett. 81, 5039-5043 (1998).}
\bibitem{rowe} M. A. Rowe, D. Kielpinski, V. Meyer, C. A. Sackett, W. M. Itano, C. Monroe and D. J. Wineland, \textit{Nature 409, 791-794 (2001).}
\bibitem{monroe} D. N. Matsukevich, P. Maunz, D. L. Moehring, S. Olmschenk, C. Monroe, \textit{Phys. Rev. Lett. 100, 150404 (2008).}
\bibitem{hoban} M. J. Hoban, E. T. Campbell, K. Loukopoulos and D. E. Browne, \textit{New J. Phys. 13 023014 (2011).}
\bibitem{berry} D. W. Berry, H. Jeong, M. Stobinska, T. C. Ralph, \textit{Phys. Rev. A 81, 012109 (2010).}
\bibitem{froissart} M. Froissart, \textit{Nuovo Cimento B64, 241 (1981).}
\bibitem{pitowsky} I. Pitowsky, Quantum Probability - Quantum Logic, \textit{Lecture Notes in Physics 321, Springer (1989).}
\bibitem{ww} R. F. Werner and M. M. Wolf, \textit{Phys. Rev. A 64, 032112 (2001).}
\bibitem{zb} M. \.{Z}ukowski, C. Brukner, \textit{Phys. Rev. Lett. 88 210401 (2002).}
\bibitem{mermin} N. D. Mermin, \textit{Phys. Rev. Lett. 65, 1838 (1990).}
\bibitem{shannon} C. Shannon, \textit{Bell System Technical Journal 28 (4): 656�715 (1949).}
\bibitem{ghz} D. M. Greenberger, M. A. Horne, A. Zeilinger, in Bell's Theorem, Quantum Theory, and Conceptions edited by M. Kafatos \textit{(Kluwer Academic, Dordrecht, 1989).}
\bibitem{anders} J. Anders, D. E. Browne, \textit{Phys. Rev. Lett. 102, 050502 (2009).}
\bibitem{mbqc} R. Raussendorf, H. J. Briegel, \textit{Phys. Rev. Lett. 86, 5188-5191 (2001).}
\bibitem{shor} P. Shor, \textit{SIAM J. Sci. Statist. Comput. 26 1481 (1997).}
\bibitem{chsh} J. F. Clauser, M.A. Horne, A. Shimony and R. A. Holt, \textit{Phys. Rev. Lett. 23, 880-884 (1969).}
\bibitem{ch} J. F. Clauser and M.A. Horne, \textit{Phys. Rev. D 10 526 (1974).}
\bibitem{tsirelson} B. S. Cirel'son, \textit{Lett. Math. Phys. 4, 93 (1980).}
\bibitem{gisin} J. Barrett, N. Gisin, \textit{Phys. Rev. Lett. 106, 100406 (2011).}
\bibitem{pr} S. Popescu, D. Rohrlich, \textit{Foundations of Physics 24, 379-385 (1994).}
\bibitem{barrett} J. Barrett, \textit{Phys. Rev. A 75, 032304 (2007).}
\bibitem{fine} A. Fine, \textit{Phys. Rev. Lett. 48, 291 (1982).}
\bibitem{supplement} See Supplementary Material.
\bibitem{thesis} M. J. Hoban, \textit{PhD Thesis (2011).}
\end{thebibliography}
\end{document}